\documentclass[conference,10pt,a4paper]{IEEEtran}

\usepackage[lmargin=0.6181102in, rmargin=0.6181102in, tmargin=0.3740157in, bmargin=0.92in]{geometry}

\usepackage{subfiles}
\usepackage{excludeonly}

\usepackage[utf8]{inputenc}
\usepackage[T1]{fontenc}
\usepackage[english]{babel}

\usepackage{mathtools}
\usepackage{amssymb} 

\usepackage{bm} 

\usepackage{pifont}

\usepackage{newtxtext,newtxmath}  

\usepackage[version=3]{mhchem} 
\usepackage{esint} 
\usepackage{dsfont} 
\usepackage[makeroom]{cancel} 
\usepackage{empheq}
\usepackage{physics}
\usepackage[binary-units=true,per-mode=symbol]{siunitx}

\usepackage{graphicx}
\usepackage{float}
\makeatletter
\let\MYcaption\@makecaption
\makeatother

\usepackage[font=footnotesize]{subcaption}

\makeatletter
\let\@makecaption\MYcaption
\makeatother

\usepackage{booktabs}
\usepackage{makecell}
\usepackage{threeparttable}
\usepackage{multirow}
 
\usepackage{enumitem}

\usepackage[dvipsnames]{xcolor}

\usepackage{tcolorbox} 

\usepackage{varioref} 
\PassOptionsToPackage{hyphens}{url}
\usepackage[hidelinks,breaklinks=true,linktoc=all]{hyperref} 
\hypersetup{colorlinks=true,
allcolors=black} 
\usepackage[nameinlink,capitalise,noabbrev]{cleveref} 
\crefformat{equation}{(#2#1#3)}
\Crefformat{equation}{Equation (#2#1#3)}
\crefrangeformat{equation}{(#3#1#4) to~(#5#2#6)}
\crefmultiformat{equation}{(#2#1#3)}%
{ and~(#2#1#3)}{, (#2#1#3)}{ and~(#2#1#3)}
\Crefmultiformat{equation}{Equations (#2#1#3)}%
{ and~(#2#1#3)}{, (#2#1#3)}{ and~(#2#1#3)}

\usepackage[noadjust]{cite} 

\usepackage{etoolbox}
\usepackage{pgf,tikz,pgfplots}
\pgfplotsset{compat=1.10}
\usetikzlibrary{shapes,arrows,calc,arrows.meta,
patterns,backgrounds,positioning,fit}
\tikzset{every picture/.style=black}
\usepackage{circuitikz}

\tikzset{three sided/.style={
        draw=none,
        append after command={
            [shorten <= -0.5\pgflinewidth]
            ([shift={(-1.5\pgflinewidth,-0.5\pgflinewidth)}]				  \tikzlastnode.north east)
        edge([shift={( 0.5\pgflinewidth,-0.5\pgflinewidth)}]				  \tikzlastnode.north west) 
            ([shift={( 0.5\pgflinewidth,-0.5\pgflinewidth)}]				  \tikzlastnode.north west)
        edge([shift={( 0.5\pgflinewidth,+0.5\pgflinewidth)}]				  \tikzlastnode.south west)            
            ([shift={( 0.5\pgflinewidth,+0.5\pgflinewidth)}]				  \tikzlastnode.south west)
        edge([shift={(-1.0\pgflinewidth,+0.5\pgflinewidth)}]				  \tikzlastnode.south east)
        }
    }
}

\usepackage{psfrag}
\usepackage{pstool} 

\makeatletter
\patchcmd{\@makecaption}
  {\scshape}
  {}
  {}
  {}
\patchcmd{\@makecaption}
  {\\}
  {:\ }
  {}
  {}
\makeatletter
\patchcmd{\@makecaption}
  {\footnotesize}{}{}{}
\patchcmd{\@makecaption}
  {\footnotesize}{}{}{}
\patchcmd{\@makecaption}
  {\footnotesize}{}{}{}
\patchcmd{\@makecaption}
  {\footnotesize}{}{}{}
\makeatother

\makeatletter
\newcommand{\getfontsize}{\f@size pt}
\makeatother

\newcommand*\ds{\displaystyle}

\newcommand*\diff{\mathop{}\!\mathrm{d}}

\newcommand*\kB{k}

\newcommand*\VB{V_{\mathrm{B}}}

\newcommand*\Wn{W_\mathrm{n}}
\newcommand*\Wp{W_\mathrm{p}}
\newcommand*\Ln{L_\mathrm{n}}
\newcommand*\Lp{L_\mathrm{p}}

\newcommand*\VDD{V_{\mathrm{DD}}}

\newcommand*\fp{f_{\mathrm{p}}}

\newcommand*\fmax{f_{\mathrm{max}}}

\newcommand{\dt}{\diff{t}}

\newcommand*\vOUTi{v_{1}}
\newcommand*\vOUTii{v_{2}}

\newcommand*\dV{\delta V}

\newcommand*\dVi{\delta V_1}
\newcommand*\dVii{\delta V_2}

\newcommand*\TTF{TTF}
\newcommand*\MTTF{MTTF}

\newcommand*\XM{X_{\mathrm{M}}}
\newcommand*\YM{Y_{\mathrm{M}}}

\renewcommand*\vv{\tilde{v}}
\newcommand*\Dvv{\Delta \tilde{v}}

\newcommand*{\EE}{\mathcal{U}}
\newcommand*\sigmaw{\sigma_{\mathrm{w}}}

\definecolor{k}{rgb}{0 0 0}
\definecolor{r}{rgb}{1 0 0}
\definecolor{g}{rgb}{0 1 0}
\definecolor{b}{rgb}{0 0 1}
\definecolor{orange}{rgb}{1,0.7,0}
\definecolor{c}{rgb}{0 1 1}
\definecolor{cc}{RGB}{64 224 208}
\definecolor{m}{rgb}{1 0 1}
\definecolor{khaki}{RGB}{128 128 0}
\definecolor{deepskyblue}{RGB}{0 191 255}
\definecolor{darkMagenta}{rgb}{0.5 0 0.5}
\definecolor{chocolateBrown}{RGB}{98 52 18}
\definecolor{lightBrown}{RGB}{189 154 122}
\definecolor{mybrown}{RGB}{127 37 0}
\definecolor{bordeaux}{RGB}{131 41 85}
\definecolor{violet}{RGB}{127 0 255}
\definecolor{myGreen}{RGB}{134,180,44}
\definecolor{gray_gate}{RGB}{211,208,205}
\definecolor{yellow_oxide}{RGB}{244,231,164}
\definecolor{color_mix}{rgb}{0.7510 0.2510 0.2510}

\definecolor{h}{rgb}{0 0 0}

\definecolor{l}{rgb}{0 0 1}

\graphicspath{{figures/}}

\usepackage{authblk}

\makeatletter
\renewcommand\AB@affilsepx{, \protect\Affilfont}
\makeatother

\renewcommand{\IEEEbibitemsep}{0.7pt plus 1pt}
\makeatletter
\IEEEtriggercmd{\reset@font\normalfont\fontsize{8pt}{10pt}\selectfont}
\makeatother
\IEEEtriggeratref{1}

\makeatletter
\def\ps@IEEEtitlepagestyle{
\let\@oddhead\@empty
\let\@evenhead\@empty
\let\@evenfoot\@empty
}
\makeatother

\begin{document}

\bstctlcite{IEEEexample:BSTcontrol} 

%
%

\title{\fontsize{14}{19}\selectfont{\textbf{
Stochastic Nonlinear Dynamical Modelling of 
SRAM Bitcells in Retention Mode
}}\vspace{-2ex}}
\author{\fontsize{12}{14}\selectfont{Léopold Van Brandt, Denis Flandre and Jean-Charles Delvenne}
\\
\fontsize{12}{14}\selectfont{ICTEAM Institute, UCLouvain, Louvain-la-Neuve, Belgium, {\tt leopold.vanbrandt@uclouvain.be}}
}

\maketitle

%
%

\section*{Abstract}

SRAM bitcells in retention mode 
behave
as autonomous stochastic nonlinear dynamical systems.
From observation of variability-aware transient noise simulations, we provide an unidimensional model, fully characterizable by conventional deterministic SPICE simulations, insightfully 
explaining 
the mechanism of intrinsic noise-induced bit flips.
\textcolor{h}{
The proposed model is exploited to, first, 
explain 
the reported inaccuracy of existing closed-form near-equilibrium formulas aimed at predicting the mean time to failure and, secondly, to propose a closer estimate 
attractive in terms of CPU time.
}\\
Keywords: Ultra-Low-Voltage SRAM, Noise-Induced Failures, SRAM Dynamic Stability, Stochastic Modelling.

\section*{Introduction}

\emph{Static Random Access Memory} (\emph{SRAM}) arrays contains hundred of thousands bitcells,
e.g $\SI{32}{\kilo\byte} = \SI{262 144}{bits}$ for an ultra low voltage (ULV) microcontroller~\cite{Bol2021_SleepRunner}, whose functionality is statistically endangered by the combined effects of \emph{process variability} and \emph{intrinsic} 
\emph{noise}~\cite{LASCAS2024}.
The effect of process variations is twofold.
On one hand, a small yet non-negligible fraction 
(e.g. that must be guaranteed 
below $\sim \SI{100}{ppm}$~\cite{Dreslinski2010}) 
of bitcells are defective as fabricated~\cite{VanBrandt2022_TCASI_SRAM}.
On the other hand, 
some surviving bitcells severely affected by variability have reduced noise immunity and are thereby prone to \emph{dynamic instability}~\cite{SSE2023}.  
Although these 
are deemed functional at time zero, the intrinsic noise is likely to induce \emph{transient failures} or \emph{bit flips} in short times unacceptable for data retention in practical 
applications~\cite{SSE2023}.

 
Simulating
bit flips in SRAM bitcells in \emph{retention} mode
requires 
transient noise analyses~\cite{Rezaei2020,SSE2023,LASCAS2024}, 
which are conventional 
time-domain simulations 
however
with independent random current noise sources added in parallel to each dissipative devices (resistors, 
MOS transistors,...).
A robust methodology, compatible
with industrial tools, was described in~\cite{SSE2023}.
Whereas transient noise simulations are suitable for nonlinear circuits operating in out-of-equilibrium, large-signal conditions such as SRAM bitcells on the verge of instability,
the 
CPU
time quickly explodes 
with the number of cases (e.g. process variations)~\cite{SSE2023,LASCAS2024}.


Reference~\cite{Rezaei2020} developed an accelerated simulator aimed at efficiently estimating the \emph{mean time to failure} ($\MTTF$)
in SRAMs though not straightforward to extend 
other memory architectures and technologies.
We also find a few analytical attempts in the literature.
In \cite{Freitas2022_reliability}, the $\MTTF$ is calculated from stochastic thermodynamic considerations, however assuming simplified transistor model and constant capacitances.
We owe the only existing closed-form formula in the electronics literature to Kish~\cite{Kish2002}, with preliminary 
attempts of application reported in~\cite{Veirano2016_journal,LASCAS2024}.
The previous work~\cite{LASCAS2024}
has shown by numerical experiments
that Kish's 
and the similar but more rigorous Nobile's formulas~\cite{Nobile1985}
lack accuracy.
Both indeed rely on 
coarse
near-equilibrium approximation around the presumed stable point of the bitcell in retention, as will be revisited in the present paper.
Ignoring technological and application aspects (notably variability),
SRAM dynamic stability has been investigated in the past within the mathematical framework of nonlinear dynamical systems~\cite{Zhang2006,Dong2008}.
Paper~\cite{Zhang2006} used simplified analytical inverter model and only considered deterministic pulse noise as disturbance, i.e. not the thermal noise of the transistors \cite{Rezaei2020,SSE2023,LASCAS2024}.



The purpose of the present work is to propose a stochastic and nonlinear dynamical model of an SRAM bitcell in retention, insightfully explaining the observed transient bit-flip mechanism.
The parameters are fully extracted from conventional deterministic SPICE simulations, compatible with industrial transistor compact models and advanced technologies. 
While pointing out the limits of existing analytical formulas at the light of the presented model, we 
open promising avenues for 
fast and accurate
semi-analytical approaches.

\section*{Modelling and Analysis}
\label{section:Modelling and Analysis}

\begin{figure}[t]
\captionsetup[subfigure]{singlelinecheck=off,justification=raggedright}
\newcommand\myfontsize{\small}
\newcommand\mytickfontsize{\footnotesize}
\captionsetup[subfigure]{skip=0pt}
\begin{subfigure}[t]{\linewidth}
\myfontsize
\centering
\subcaption{}
\label{fig_SRAM}
\begin{circuitikz}[american voltages, transform shape, line cap=rect, nodes={line cap=butt},scale=1]
\draw
(0,0) node[not port,color=black] (inv) {}
++ (0,-2) node[not port,xscale=-1,color=black] (inv2) {};
\draw[black]
(inv.out) -- ++(2,0) node[label={[font=\myfontsize,color=red]above:$\vOUTi(t)$},color=red] (vout) {$\bullet$}
-- ++(0,-2) node[] (vin2) {};
\draw[black]
(inv2.in) ++ (1.5,0) node[] (vinn2) {}
(vinn2) to[V=$\delta V_2$,invert,scale=1,color=black] (inv2.in);
\draw[black]
(vinn2.center) -- (vin2.center);
\draw[black]
(inv.in) ++ (-2,0) node[] (vin) {}
-- ++(0,-2) node[label={[font=\myfontsize,color=gray]below:$\vOUTii(t)$},color=gray] (vout2) {$\bullet$}
-- (inv2.out)
(inv.in) ++ (-1.5,0) node[] (vinn) {}
(vinn) to[V=$\delta V_1$,invert,scale=1] (inv.in)
(vin.center) -- (vinn.center);
\draw
(-3.5,-1) node[nmos,rotate = -90,yscale=-1] (nMOSl) {}
(nMOSl.G) to[short,-o] ++(0,0) node[label={[font=\scriptsize,color=black,yshift=+0.0cm]above:$V_{\mathrm{WL}} = 0$}] (nGl) {}
(vout2) ++ (0,+1) node[] (vout2m) {}
(nMOSl.S) -- (vout2m.center)
(nMOSl.D) -- ++ (0,2) node[label={[font=\scriptsize]right:$V_{\mathrm{BL}} = \VDD$}] (VBLl) {}
(nMOSl.D) -- ++ (0,-1.5) 
;
\draw
(+3.5,-1) node[nmos,rotate = -90,yscale=-1] (nMOSr) {}
(nMOSr.G) to[short,-o] ++(0,0) node[label={[font=\scriptsize,color=black,yshift=+0.0cm]above:$V_{\mathrm{WL}} = 0$}] (nGl) {}
(vout) ++ (0,-1) node[] (vout1m) {}
(nMOSr.D) -- (vout1m.center)
(nMOSr.S) -- ++ (0,2) node[label={[font=\scriptsize]left:$V_{\mathrm{BL}} = \VDD$}] (VBLr) {}
(nMOSr.S) -- ++ (0,-1.5) 
;
\draw
(vinn)++(0.0,0) node[coordinate] (l1) {}
(inv)++(0,1) node[coordinate] (t1) {}
(inv)++(0,-0.75) node[coordinate] (b1) {}
node[fit={(l1)(inv.out)(t1)(b1)},draw, rectangle, dashed, black,
label={[font=\myfontsize,yshift=-0.50cm]right:Inverter 1}
, inner sep=0pt] {};
\draw
(vinn2)++(0.0,0) node[coordinate] (r2) {}
(inv2)++(0,0.75) node[coordinate] (t2) {}
(inv2)++(0,-1) node[coordinate] (b2) {}
node[fit={(inv2.out)(r2)(t2)(b2)},draw, rectangle, dashed, black,
label={[font=\myfontsize,yshift=+0.50cm]left:Inverter 2}
, inner sep=0pt] {};
\end{circuitikz}
\end{subfigure}
\begin{subfigure}[t]{\linewidth}
\myfontsize
\centering
\vspace{-1mm}
\subcaption{}
\label{fig_bit_flip}
\psfragscanon
\psfrag{v(t) [mV]}[cc][cc]{$v(t)$}
\psfrag{474}[cc][cc]{\mytickfontsize$474$}
\psfrag{t [us]}[cc][cc]{$t \, [\si{\micro\second}]$}
\psfrag{475}[cc][cc]{\mytickfontsize$475$}
\psfrag{TTF}[cc][cc]{\mytickfontsize$\TTF$}
\psfrag{Y0}[cc][cc]{\color{b}$Y_0$}
\psfrag{YM}[cc][cc]{\color{b}$\YM$}
\psfrag{XM}[bc][cc]{\color{b}$\XM$}
\psfrag{X0}[cc][cc]{\color{b}$X_0$}
\psfrag{X1}[cl][cl]{\color{b}$X_1$}
\psfrag{Y1}[cl][cl]{\color{b}$Y_1$}
\psfrag{vOUT2(t)}[bl][bl]{$\textcolor{gray}{\vOUTii(t)}$}
\psfrag{vOUT1(t)}[tl][tl]{$\textcolor{r}{\vOUTi(t)}$}
\includegraphics[scale=1]{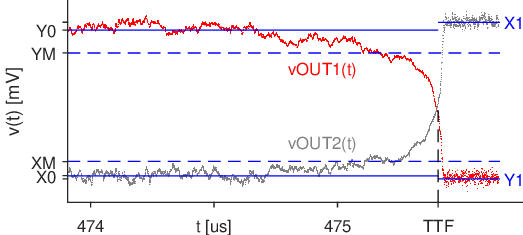}
\end{subfigure}
\captionsetup[subfigure]{skip=0pt}
\begin{subfigure}[t]{\linewidth}
\myfontsize
\centering
\vspace{-1mm}
\subcaption{}
\label{fig_state_space}
\newcommand\mycolor{b}
\psfragscanon
\psfrag{vIN1 = vOUT2 [mV]}[cc][cc]{$\textcolor{gray}{\vOUTii(t)} \, [\si{\milli\volt}]$}
\psfrag{vOUT1 = vIN2 [mV]}[cc][cc]{$\textcolor{r}{\vOUTi(t)} \, [\si{\milli\volt}]$}
\psfrag{0}[cc][cc]{\mytickfontsize$0$}
\psfrag{50}[cc][cc]{\mytickfontsize$50$}
\psfrag{100}[cc][cc]{\mytickfontsize$100$}
\psfrag{150}[cc][cc]{\mytickfontsize$150$}
\psfrag{200}[cc][cc]{\mytickfontsize$200$}
\psfrag{(X0,Y0)}[cr][cr]{\footnotesize$\textcolor{\mycolor}{(X_0,Y_0)}$}
\psfrag{(XM,YM)}[cl][cl]{\footnotesize$\textcolor{\mycolor}{(\XM,\YM)}$}
\psfrag{(X1,Y1)}[br][br]{\footnotesize$\textcolor{\mycolor}{(X_1,Y_1)}$}
\psfrag{Nominal}[tl][tl]{\color{myGreen}
Nominal VTCs ($\dVi = -\dVii = 0$)
}
\psfrag{dV1 = -dV2 - 55 mV}[bl][bl]{\color{\mycolor}$\dVi = -\dVii = \SI{55}{\milli\volt}$}
\psfrag{(vOUT2(t),vOUT1(t))}[bl][bl]{\color{color_mix}$(\vOUTii(t),\vOUTi(t))$}
\psfrag{vv}[tc][tc]{\color{violet} $\vv$}
\includegraphics[scale=1]{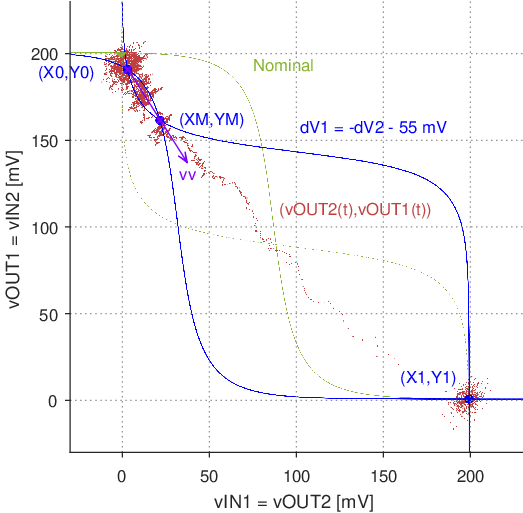}
\end{subfigure}
\caption{
\subref{fig_SRAM} SRAM bitcell in retention mode,
with series-voltage sources modelling process variations.
Illustrated below: 
$\dVi = -\dVii = \SI{55}{\milli\volt}$.
\newline
\subref{fig_bit_flip}
Transient simulation of a noise-induced bit flip in the 6T SRAM bitcell.
\newline
\subref{fig_state_space}
State trajectory of the bit flip of \subref{fig_bit_flip} in the state space. 
The nominal and modified VTCs of the inverters are shown in blue and green, respectively.
The preferential reaction coordinate $\vv$ is shown in violet.
\newline
Illustrated case: $\SI{28}{\nano\meter}$ FD-SOI Single-P-Well (SPW) SRAM cell (inverters made of RVT nMOS and LVT pMOS; RVT nMOS access transistors)
of minimal transistor dimensions \mbox{$\Ln = \Lp = \SI{30}{\nano\meter}$} and \mbox{$\Wn = \Wp = \SI{80}{\nano\meter}$}, 
\mbox{SPW bias $\VB = 0$}, 
operating at \mbox{$\VDD = \SI{200}{\milli\volt}$} 
and $T = \SI{300}{\kelvin}$.
\newline
Bandwidth of 
the 
generated 
noise: 
$\fmax = \SI{1}{\giga\hertz}$
($\dt = \SI{500}{\pico\second}$)~\cite{LASCAS2024}. 
\vspace{-3mm}
}
\label{fig_bit_flip_state_space}
\end{figure}

In Six-Transistor (6T) and other ULV SRAM bitcell architectures, 
data \emph{retention} is ensured by a cross-coupled inverter pair (\cref{fig_SRAM}) implementing a feedback loop 
counteracting moderate disturbances~\cite{LASCAS2024}.
We simulate the noise in the variability-aware setup proposed in~\cite{LASCAS2024}.
Following~\cite{VanBrandt2022_TCASI_SRAM}, process variations are introduced as series-voltage sources $\dVi$ and $\dVii$ applied at the input of the inverters (\cref{fig_SRAM}). 
Their effect is incorporated in the \emph{modified} voltage transfer characteristics (VTCs, in blue in \cref{fig_state_space}), shifted from the nominal VTCs (in green).
Without loss of generality, we focus the 
noise 
analyses
on the special case $\dVi = -\dVii$ corresponding to the worse-case scenario where both inverters are adversely affected~\cite{LASCAS2024}.

The two output node voltages are denoted $\vOUTi(t)$ and $\vOUTii(t)$, respectively.
If we adopt the convention $\dVi = -\dVii > 0$, the threatened memory state is $(\vOUTii,\vOUTi) = (X_0,Y_0)$ (see \cref{fig_state_space})
due to the degraded noise margins.

\subsection{Observation and Analysis of the Bit-Flip Mechanism}
\label{subsection:Observation and Analysis of the Bit-Flip Mechanism}


\Cref{fig_bit_flip} presents one typical simulation of intrinsic noise-induced bit flip~\cite{LASCAS2024}.
The underlying mechanism is more easily understood with the \emph{state space} representation~\cite{Zhang2006,Dong2008} of \cref{fig_state_space}, where the \emph{state vectors} $(\vOUTii(t),\vOUTi(t))$ are plotted at various times to yield the \emph{state trajectory}.
The VTCs of the affected bitcell are also represented to locate the two \emph{stable} 
\emph{equilibrium points} $(X_0,Y_0)$ and $(X_1,Y_1)$, which slightly deviate from the 
nominal
$(0,\VDD)$ and $(\VDD,0)$ due to process variations 
and to 
highlight
the out-of-equilibrium behaviour of the 
SRAM during the transient bit flip.

The \emph{unstable} equilibrium point $(\XM,\YM)$ is precisely located on the \emph{stability boundary} (or \emph{separatrix}~\cite{Dong2008}) separating the two stability regions.
As long as $(\vOUTii(t),\vOUTi(t))$ lies before $(\XM,\YM)$, 
the natural dynamics of the SRAM bitcell tends to bring it to back to its initial logic state $(X_0,Y_0)$ in absence of continuous disturbance.
A state flip occurs when $(\vOUTii(t),\vOUTi(t))$ crosses
$(\XM,\YM)$ due to 
simultaneous large voltage noise fluctuations (\cref{fig_bit_flip}).
Once $(\vOUTii(t),\vOUTi(t))$ has fallen in the region of attraction of the other equilibrium, $(X_1,Y_1)$, the two cross-coupled inverters enter in positive feedback loop that quickly completes the bit flip.
We define the random $\TTF$ as the time when $\vOUTii(t)$ and $\vOUTi(t)$ cross (\cref{fig_bit_flip}).
\subsection{Stochastic Nonlinear Dynamical Model}
\label{subsection:Stochastic Nonlinear Dynamical Model}

\begin{figure}[]
\captionsetup[subfigure]{singlelinecheck=off,justification=raggedright}
\newcommand\myfontsize{\footnotesize}
\newcommand\mytickfontsize{\scriptsize}
\captionsetup[subfigure]{skip=0pt}
\begin{subfigure}[t]{\linewidth}
\myfontsize
\centering
\psfragscanon
\psfrag{vv [mV]}[cc][cc]{\myfontsize$\textcolor{violet}{\vv} \, [\si{\milli\volt}]$}
\psfrag{dvv/dt [mV/us]}[cc][cc]{\myfontsize$h(\vv) \, [\si{\milli\volt\per\micro\second}]$}
\psfrag{-50}[cc][cc]{\mytickfontsize$-50$}
\psfrag{-100}[cc][cc]{\mytickfontsize$-100$}
\psfrag{0}[cc][cc]{\mytickfontsize$0$}
\psfrag{5}[cc][cc]{\mytickfontsize$5$}
\psfrag{10}[cc][cc]{\mytickfontsize$10$}
\psfrag{15}[cc][cc]{\mytickfontsize$15$}
\psfrag{20}[cc][cc]{\mytickfontsize$20$}
\psfrag{25}[cc][cc]{\mytickfontsize$25$}
\psfrag{30}[cc][cc]{\mytickfontsize$30$}
\psfrag{35}[cc][cc]{\mytickfontsize$35$}
\psfrag{40}[cc][cc]{\mytickfontsize$40$}
\psfrag{Linear}[cl][cl]{\footnotesize\color{r}Near-equilibrium linearisation: $\ds -\frac{1}{\tau} \vv$}
\includegraphics[scale=1]{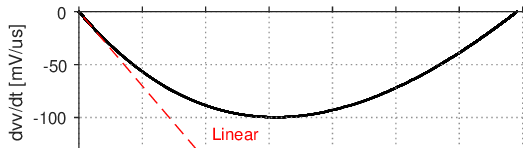}
\vspace{-2.5cm}
\vspace{-\baselineskip}
\subcaption{}
\label{fig_h}
\end{subfigure}

\vspace{+3mm}

\captionsetup[subfigure]{skip=0pt}
\begin{subfigure}[t]{\linewidth}
\myfontsize
\centering
\psfragscanon
\psfrag{vv [mV]}[cc][cc]{\myfontsize$\textcolor{violet}{\vv} \, [\si{\milli\volt}]$}
\psfrag{E(vv) [V2/s]}[cc][cc]{\myfontsize$\EE(\vv) \, [\si{\square\volt\per\second}]$}
\psfrag{0}[cc][cc]{\mytickfontsize$0$}
\psfrag{5}[cc][cc]{\mytickfontsize$5$}
\psfrag{10}[cc][cc]{\mytickfontsize$10$}
\psfrag{15}[cc][cc]{\mytickfontsize$15$}
\psfrag{20}[cc][cc]{\mytickfontsize$20$}
\psfrag{25}[cc][cc]{\mytickfontsize$25$}
\psfrag{30}[cc][cc]{\mytickfontsize$30$}
\psfrag{35}[cc][cc]{\mytickfontsize$35$}
\psfrag{40}[cc][cc]{\mytickfontsize$40$}
\psfrag{1000}[cc][cc]{\mytickfontsize$1000$}
\psfrag{2000}[cc][cc]{\mytickfontsize$2000$}
\psfrag{3000}[cc][cc]{\mytickfontsize$3000$}
\psfrag{Parabolic}[cl][cl]{\footnotesize\color{r}Parabolic potential: $\ds \frac{1}{2}\frac{1}{\tau} \vv^2$}
\includegraphics[scale=1]{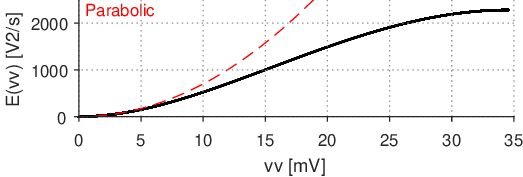}
\vspace{-3cm}
\vspace{-\baselineskip}
\vspace{-1mm}
\subcaption{}
\label{fig_E}
\end{subfigure}
\caption{
Extraction of the deterministic drift term in \eqref{eq:dvv/dt}.
Near-equilibrium approximations inherent to Kish's formula are shown in red.
\subref{fig_h}
$h(\vv)$.
\subref{fig_E}
Quasi potential $\EE(\vv) = - \int_{0}^{\vv} h(\vv') \,\diff{\vv'}$.
Illustrated case: same as \cref{fig_bit_flip_state_space}.
\vspace{-3mm}
}
\label{fig_h_E}
\end{figure}

Upon observation of the experience in \cref{fig_bit_flip,fig_state_space} (other trajectories simulated for other cases showed similar behaviour), it seems reasonable to assume that a bit flip occurs according to the \emph{preferential direction} given by the straight line connecting the two nearby points $(X_0,Y_0)$ and $(\XM,\YM)$.
This justifies the introduction of the unidimensional coordinate $\vv$ (shown in violet in \cref{fig_state_space}) constructed by affine transformation (orthogonal projection) of $(\vOUTii,\vOUTi)$: 
\begin{equation}
\label{eq:vv}
\vv \equiv
\frac{\XM - X_0}{\Dvv} \cdot \vOUTii
+
\frac{\YM - Y_0}{\Dvv} \cdot \vOUTi
\text{,}
\end{equation}
where $\YM - Y_0 < 0$ and 
$\Dvv \equiv \sqrt{(\XM-X_0)^2+(Y_0-\YM)^2}$.

{\color{h}
We assume a first-order dynamics:
\vspace{-2mm}
\begin{equation}
\label{eq:dvv/dt}
\diff{\vv}/\diff{t}
= h(\vv) + \sigmaw w(t)
\text{,}
\end{equation}
which 
may be thought as formally describing a \emph{nonlinear} $RC$ circuit, i.e. an overdamped 
single-state
system.
The function $h(\vv)$ absorbs the nonlinearities of both the resistive and capacitive components of the SRAM.
}
It can be cheaply and unambiguously extracted from a noiseless transient simulation:
1) starting from an initial condition $(\vOUTii(0),\vOUTi(0)) = (\XM - \epsilon,\YM + \epsilon)$; the deterministic evolution of $(\vOUTii(t),\vOUTi(t))$ toward ($X_0,Y_0)$ is recorded; 2) $\vv(t)$ is calculated through \eqref{eq:vv}; 3) 
$\diff{\vv}/\diff{t}$
is computed numerically and mapped to $\vv(t)$ to provide $h(\vv)$ between $0$ and $\Dvv$ (depicted in \cref{fig_h}).

{\color{h}

In \eqref{eq:dvv/dt}, $\sigmaw w(t)$ is the thermal noise term, $\sigmaw^2$ being the variance per unit time (i.e. in $\si{\square\volt\per\second}$) and 
$w(t)$ the white noise process of unit variance
(in
$\si{\second^{-1/2}}$).
\Cref{eq:dvv/dt} may also be regarded as a nonlinear \emph{drift} ($h(\vv)$) - \emph{diffusion} ($\sigmaw w(t)$) dynamics,
considering that
the drift term 
derives from a scalar \emph{quasi potential} $\EE(\vv)$ (in $\si{\square\volt\per\second}$), $h(\vv) \equiv -\diff{\EE}/\diff{\vv}$, as represented in \cref{fig_E}.
The stable point $\vv = 0$ corresponds to the valley of $\EE(\vv)$, whereas the 
hill
characterizes the unstable point at $\vv = \Dvv$. 

Thereafter, we assume that $\sigmaw^2$ does \emph{not} vary with $\vv$, and equals to its equilibrium value at $\vv = 0$.
Only at the stable equilibrium point, the SRAM dynamics can be formally assimilated to a linear $RC$ circuit and $\sigma_{\vv}^2$, the time-independent variance of $\vv(t)$, can be soundly defined and extracted by combining noise \texttt{AC} simulations with relation \eqref{eq:vv} 
(see ~\cite[(2)]{LASCAS2024}).
For an $RC$ circuit, we have $\tau = RC$, $\sigma_{\vv}^2 = \kB T/C$ and $\sigmaw^2 = 2\kB T/(R C^2)$
(Johnson-Nyquist formula). 
These combine into
$\sigmaw^2 = 2 \sigma_{\vv}^2/\tau$, where $\tau$ is the time constant of the linearised system (extraction illustrated in \cref{fig_h}).

}

\section*{Discussion of Existing Analytical Predictions}
\label{Discussion of Existing Analytical Predictions}

\begin{figure}[]
\newcommand\myfontsize{\small}
\newcommand\mytickfontsize{\footnotesize}
\myfontsize
\psfragscanon
\psfrag{dV1 = -dV2 [mV]}[cc][cc]{\myfontsize$\dVi = -\dVii \, [\si{\milli\volt}]$}
\psfrag{MTTF [s]}[cc][cc]{\myfontsize$MTTF \, [\si{\second}]$}
\psfrag{55}[cc][cc]{\mytickfontsize$55$}
\psfrag{56}[cc][cc]{\mytickfontsize$56$}
\psfrag{57}[cc][cc]{\mytickfontsize$57$}
\psfrag{58}[cc][cc]{\mytickfontsize$58$}
\psfrag{59}[cc][cc]{\mytickfontsize$59$}
\psfrag{60}[cc][cc]{\mytickfontsize$60$}
\psfrag{e0}[cr][cr]{\mytickfontsize$1$}
\psfrag{em1}[cr][cr]{\mytickfontsize$10^{-1}$}
\psfrag{em2}[cr][cr]{\mytickfontsize$10^{-2}$}
\psfrag{em3}[cr][cr]{\mytickfontsize$10^{-3}$}
\psfrag{em4}[cr][cr]{\mytickfontsize$10^{-4}$}
\psfrag{em5}[cr][cr]{\mytickfontsize$10^{-5}$}
\psfrag{em6}[cr][cr]{\mytickfontsize$10^{-6}$}
\psfrag{em7}[cr][cr]{\mytickfontsize$10^{-7}$}
\psfrag{em8}[cr][cr]{\mytickfontsize$10^{-8}$}
\psfrag{em9}[cr][cr]{\mytickfontsize$10^{-9}$}
\psfrag{NOISETRAN}[cl][cl]{\color{b}Transient noise simulations}
\psfrag{Kish}[cl][cl]{\color{r}Kish \eqref{eq:MTTF Kish}~\cite{Kish2002}}
\psfrag{Nobile}[cl][cl]{\color{orange}Nobile \cite{Nobile1985}}
\psfrag{SDE}[cl][cl]{\color{violet}Model \eqref{eq:dvv/dt}}
\includegraphics[scale=1]{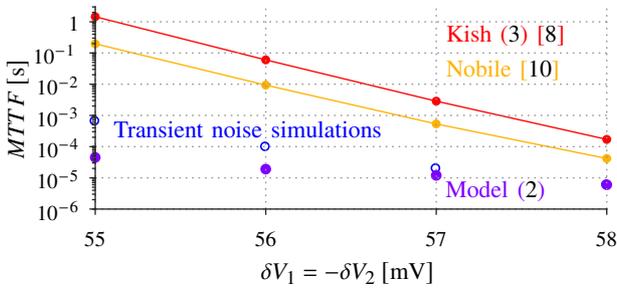}
\caption{
Comparison between the $\MTTF$ estimated empirically from transient noise simulations (\num{100} like \cref{fig_bit_flip} for each $\dV$ case \cite{LASCAS2024}), from the model \eqref{eq:dvv/dt}, and predictions of analytical near-equilibrium formulas.
\vspace{-3mm}
}
\label{fig_MTTF}
\end{figure}

Reference values for the $\MTTF$ were obtained by averaging \num{100} $\TTF$ extracted from SPICE transient noise simulations like \cref{fig_bit_flip}~\cite{LASCAS2024}, for 
several 
process variations $\dVi = -\dVii = 55$ to $\SI{58}{\milli\volt}$
(\cref{fig_MTTF}).
The total CPU time reaches
 no less than 
a few \emph{days}~\cite{LASCAS2024}.
The model \eqref{eq:dvv/dt} is a stochastic differential equation for which trajectories can be efficiently simulated in a Monte-Carlo fashion by resorting to a dedicated numerical integration schema like the \emph{Euler–Maruyama method}~\cite{Higham2001}.
For each $\vv(t)$ 
sample paths,
one $\TTF$ is obtained by recording the first time to reach $\Dvv$, multiplied by \num{2} since at the top of the hill (\cref{fig_E}) the state can still flip left or right with probability $1/2$.
The $\MTTF$ thus estimated are shown in violet in \cref{fig_MTTF}.
Very good agreement can be observed, excepted for $\dV = \SI{55}{\milli\volt}$. The distance $\Dvv$ is larger for this case, and the discrepancy is likely to be attributed to the constant-$\sigmaw$ assumption which, yet convenient, is physically questionable.
We can appreciate a CPU time reduced to barely a few \emph{minutes}.

We now review existing closed-form formulas from the literature.
Kish proposed a simplified Rice formula for the mean frequency of crossing ($1/\MTTF$) a given threshold voltage ($\Dvv$) by a band-limited Gaussian white noise process ($\vv(t)$)~\cite[(9)]{Kish2002}:

\vspace{-2mm}
{\color{h}
{\small
\begin{equation}
\label{eq:MTTF Kish}
\frac{1}{\MTTF} = \frac{2}{\sqrt{3}} \, \exp \bigg( - \frac{1}{2} \Big(\frac{\Dvv}{\sigma_{\vv}}\Big)^2 \bigg)
\, \fp
\propto \exp \bigg(  
- \frac{\Dvv^2/(2\tau)}{\sigma_{\vv}^2/\tau}
\bigg) \, \frac{1}{\tau}
\text{.}
\end{equation}
}
$\fp = 1/(2\pi\tau)$ is the $\vv$ noise bandwidth, defined at \mbox{$\vv = 0$}.
Despite lack of detailed proof, we understand \eqref{eq:MTTF Kish} as linearising the true SRAM dynamics around equilibrium to reduce it to an $RC$-like circuit with Gaussian voltage distribution.
$\Dvv^2/(2\tau)$ is the value of the quasi potential barrier evaluated at the unstable point ($\vv = \Dvv$) under the harmonic (parabolic) potential approximation (\cref{fig_E}).
\Cref{eq:MTTF Kish} 
finally highlights
a Maxwell-Boltzmann probability: 
\newline
\mbox{$1/\MTTF \propto \exp(-\Delta E/\kB T) / \tau$, with $\Delta E = C \Dvv^2 /2$.}

A similar but more mathematically rigorous $\MTTF$ formula 
can be obtained from Nobile's mean \emph{first passage time} of an Ornstein-Uhlenbeck process~\cite[(6a) multiplied by \num{2}]{Nobile1985}\cite[(5)]{LASCAS2024},
which is is defined by a linearised drift term $h(\vv) \approx - \vv/\tau$ (\cref{fig_h})~\cite{LASCAS2024}.
Consequently, Nobile relies on exactly the \emph{same} near-equilibrium approximation as Kish.

}

The above analytical predictions
were added in \cref{fig_MTTF}, respectively in red and orange, for comparison.
We emphasize that both formula totally fail in accurately predicting the $\MTTF$, the discrepancy increasing to more than two orders of magnitude for larger $\MTTF$ (more moderate variability conditions).
At the light of the model \eqref{eq:dvv/dt}
we attribute the reported inaccuracy to the near-equilibrium approximation on which the formulas are based, inappropriate to capture the nonlinear SRAM dynamics (\cref{fig_h_E}).
Through the parabolic approximation, Kish overestimates the quasi potential barrier that the process $\vv(t)$ must cross.
For the $\dVi = -\dVii =  \SI{55}{\milli\volt}$ case illustrated in \cref{fig_E}, we find that $\Dvv^2/(2\tau)$ exceeds the actual $\EE(\Dvv)$ by a factor close to \num{4}.
The Boltzmann probability $\propto \exp(-\Delta E/\kB T)$ is therefore strongly underestimated.
Although this partially explains the significant overestimation of the $\MTTF$, we must mention that the $\tau$ present in 
\eqref{eq:MTTF Kish} 
is also a quantity defined and extracted at equilibrium, hence likely to be a source of discrepancy.
{\color{h}
Remedy these existing formulas would require more general out-of-equilibrium concepts that would be consistent with the stochastic nonlinear dynamics modelled by \eqref{eq:dvv/dt} and valid from $\vv = 0$ to $\vv = \Dvv$, 
like
the Monte-Carlo trials proposed in this work.
A 
subsequent
promising approach will be to \emph{calculate} the $\MTTF$ piecewise, over small interval over which the dynamics can be \emph{locally} approximated. 
}

\section*{Conclusions}
\label{section:Conclusions}

In 
ULV SRAM design, 
bitcells with lowered noise immunity due to process variations are prone to dynamic instability.
Fast and accurate prediction of the time to 
noise-induced transient failures,
compatible with industrial models and simulation tools, 
remains a challenge.
We have proposed a stochastic nonlinear dynamical model of SRAM bitcells in retention mode.
Existing closed-form formulas are based on a near-equilibrium approximation, interpreted either as a parabolic approximation of the quasi potential barrier or, equivalently, as a linearised drift term.
We have shown that they struggle to accurately predicted the $\MTTF$ extracted from computational intensive SPICE transient noise simulations.
\textcolor{h}{
We expect from preliminary Monte-Carlo trials exploiting our model that 
a semi-analytical formulation accommodating the true SRAM dynamics and
involving parameters that can be cheaply extracted from deterministic SPICE simulations is the most promising avenue for future work on that topic.
}
The effect of random telegraph noise also needs to be investigated.

%
%

\bibliographystyle{IEEEtran}
\bibliography{IEEEabrv,bib_short}

\end{document}